\documentclass[journal]{IEEEtran}
\usepackage{epsfig,latexsym,amssymb,amsmath,graphics,color}
\usepackage{graphicx,subfigure}
\newtheorem{prop}{Proposition}

\newtheorem{cor}{Corollary}

\newtheorem{lm}{Lemma}

\newtheorem{thm}{Theorem}

\newcommand{\be}{\begin{eqnarray}}
\newcommand{\ee}{\end{eqnarray}}
\newcommand{\benn}{\begin{eqnarray*}}
\newcommand{\eenn}{\end{eqnarray*}}
\def\IR{\rm I \kern-0.20em R}

\newcommand{\bthm}{\begin{thm}}
\newcommand{\ethm}{\end{thm}}

\newcommand{\bcor}{\begin{cor}}
\newcommand{\ecor}{\end{cor}}
\newcommand{\bprop}{\begin{prop}}
\newcommand{\eprop}{\end{prop}}
\newcommand{\blm}{\begin{lm}}
\newcommand{\elm}{\end{lm}}
\newcommand{\beq}{\begin{equation}}
\newcommand{\eeq}{\end{equation}}
\newcommand{\ber}{\begin{eqnarray}}
\newcommand{\eer}{\end{eqnarray}}

\newcommand{\bproof}{\begin{proof}}
\newcommand{\eproof}{\end{proof}}

\newcommand{\bit}{\begin{itemize}}
\newcommand{\eit}{\end{itemize}}
\newcommand{\ben}{\begin{enumerate}}
\newcommand{\een}{\end{enumerate}}
\newcommand{\bdesc}{\begin{description}}
\newcommand{\edesc}{\end{description}}
\newcommand{\beqarrn}{\begin{eqnarray*}}
\newcommand{\eeqarrn}{\end{eqnarray*}}
\newcommand{\bproofof}{\begin{proofof}}
\newcommand{\eproofof}{\end{proofof}}
\newenvironment{rem}{\begin{trivlist}\item[]{\bf
Remark:}\hspace{4mm}}{\end{trivlist}}
\newcommand{\brem}{\begin{rem}}
\newcommand{\erem}{\end{rem}}
\newenvironment{rems}{\begin{trivlist}\item[]{\bf
Remarks}\begin{itemize}}{\end{itemize}\end{trivlist}}
\newcommand{\brems}{\begin{rems}}
\newcommand{\erems}{\end{rems}}
\newtheorem{fact}{Fact}
\newcommand{\bfact}{\begin{fact}}
\newcommand{\efact}{\end{fact}}
\newtheorem{examp}{Example}
\newcommand{\bexamp}{\begin{examp}\rm}
\newcommand{\eexamp}{\end{examp}}
\newtheorem{defn}{Definition}
\newcommand{\bdefn}{\begin{defn}\rm}
\newcommand{\edefn}{\end{defn}}

\newtheorem{alg}{Algorithm}
\newcommand{\balg}{\begin{alg}}
\newcommand{\ealg}{\end{alg}}

\newtheorem{prob}{Problem}
\newcommand{\bprob}{\begin{prob}}
\newcommand{\eprob}{\end{prob}}

\newcommand{\bvtm}{\begin{verbatim}}
\newcommand{\bfig}{\begin{figure}}
\newcommand{\efig}{\end{figure}}
\newcommand{\bcen}{\begin{center}}
\newcommand{\ecen}{\end{center}}

\def\beqa{\begin{eqnarray}}
\def\eeqa{\end{eqnarray}}

\long\def\comment#1{}

\def \n2{{N_0 \over 2}}

\def \h5{\hspace{0.5in}}

\begin{document}

\title{Phototransistor-like Light Controllable IoT Sensor based on Series-connected RGB LEDs}
\author{
Shangbin Li, Shuang Liang, and Zhengyuan Xu
\thanks{This work was supported by Key Program of National Natural Science Foundation of China (Grant No. 61631018), National Natural Science Foundation of China (Grant No. 61501420), Key Research Program of Frontier Sciences of CAS (Grant No. QYZDY-SSW-JSC003).
}
\thanks{S. Li, S. Liang and Z. Xu are with Key Laboratory of Wireless-Optical Communications, Chinese Academy of Sciences, University of Science and Technology of China, Hefei, Anhui 230027, China. Email: \{shbli, xuzy\}@ustc.edu.cn.}}

\maketitle
\thispagestyle{empty}
\pagestyle{empty}

\begin{abstract}
An IoT optical sensor based on the series-connected RGB LEDs is designed, which exhibits the light-controllable optical-to-electrical response like a phototransistor. The IoT sensor has the maximal AC and DC responsivities to the violet light mixed by blue and red light. Its responsivity to the blue light is programmable by the impinging red or green light. A theoretical model based on the light-dependent impedance is developed to interpret its novel optoelectronic response. Such IoT sensor can simultaneously serve as the transmitter and the receiver in the IoT optical communication network, thus significantly reduces the system complexity.
\end{abstract}
\renewcommand{\IEEEkeywordsname}{Index Terms}

\vspace*{-0.6cm}
\noindent
\section{Introduction}

Visible light communication (VLC) based internet of things (IoT) sensors and applications have attracted much attention \cite{26,27,28}. The light-emitting diodes (LEDs) are typically used as emitters in the VLC systems \cite{1,2,3}. In addition to electrical-to-optical (EO) conversion, the LEDs can also serve as optical detectors via the optical-to-electrical (OE) conversion \cite{5,6,7}. The LED-LED VLC systems with time-division half duplex \cite{11,12,13} and full duplex \cite{19} have been investigated and demonstrated. The alternate current (AC) impedance of the LED receiver can be controlled by the injected light \cite{20}. InGaN-based or InGaAsP-based phototransistors have also been studied \cite{22,23}. The former is advantageous in low dark current and high responsivity of the white-LED based VLC receivers. Recently, a novel all-optical-input transistor based on Ag/TiO2 is demonstrated \cite{29}, and a prototype CMOS active sensor simultaneously imaging and energy harvesting is presented \cite{25}.

Here, we firstly design an optical IoT sensor consisting of series-connected RGB LEDs, and experimentally demonstrate its light controllable phototransistor-like characteristics. The intriguing phenomenon of the series-connected RGB LED sensor (abbreviated RGB sensor throughout this paper) is its light-programmable OE response. The RGB sensor responds to the AC light signal and the direct current (DC) light signal differently, depending on the color of input AC or DC light signal or their color mixture. For example, the response to the AC light signal can be controlled by the input DC light with a different color or a mixture of colors. Its red and green AC signal responses can be effectively suppressed by the DC input of the green and red light, respectively, while its blue AC signal response can be enhanced by the red and green DC input light. Thus, the RGB sensor is also suitable for the yellow phosphor coated white LED based VLC applications due to its unique behavior. The low speed yellow light signal of the phosphor does not induce the significant interference but enhances the response of the RGB sensor to the high speed blue light signal.

\section{RGB sensor model and experimental scheme}

Figure \ref{fig1}(a) depicts the experimental scheme for the optoelectronic response of the RGB sensor consisting of the series-connected AlInGaP red LED, InGaN green LED and GaN blue LED, where the LEDs belong to the LumiLEDs rebel series. Other commercial RGB LEDs can also be utilized, and the results will be discussed elsewhere. In the experiments, all  LEDs in the RGB sensor are uniformly illuminated no matter what color of the injected light. The output of the RGB sensor is connected to the Keysight MSOX6004A oscilloscope with the input load $1$ M$\Omega$. In Fig. \ref{fig1}(b), the measured emission spectrum and relative optoelectronic response spectral distributions of the RGB LEDs are presented \cite{19}.

\vspace*{-0.2cm}
\begin{figure}[ht]
\centering
\subfigure[]{\includegraphics[width=0.5\columnwidth,angle=-0]{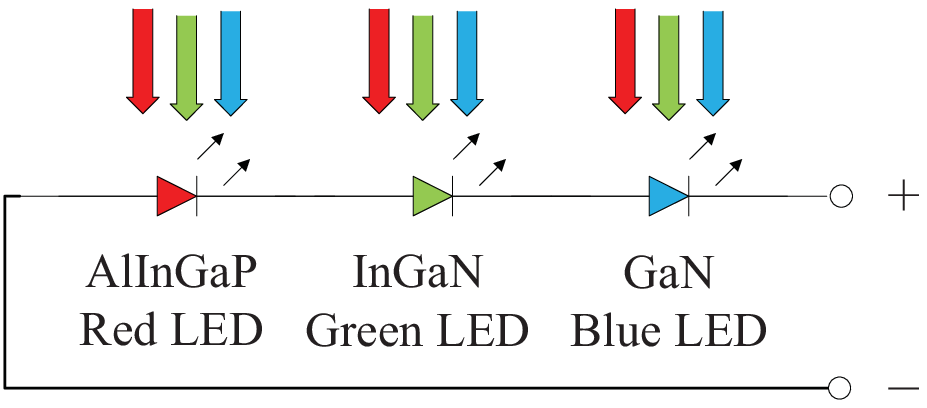}}
\subfigure[]{\includegraphics[width=0.7\columnwidth,angle=-0]{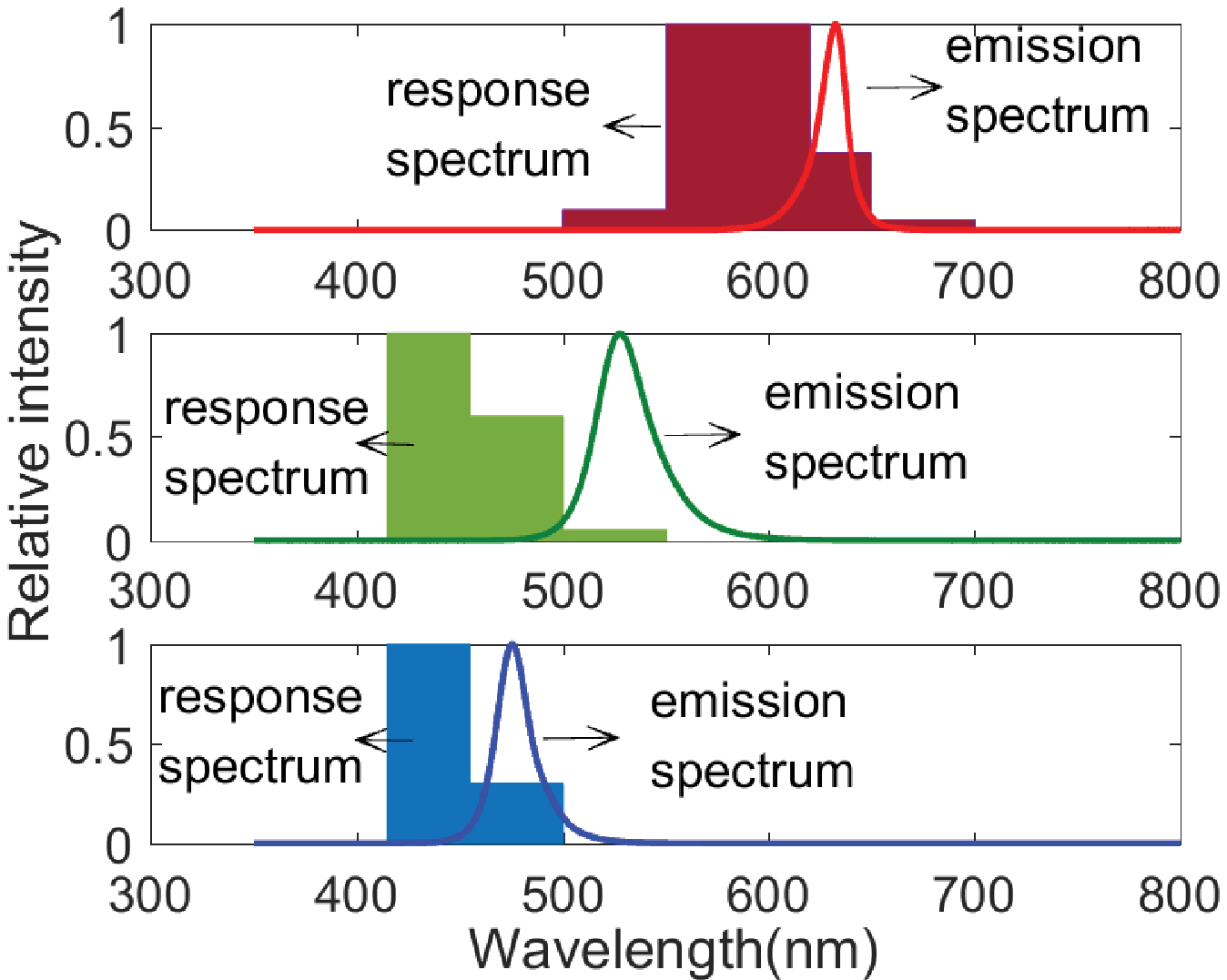}}
\caption{(a) Experimental scheme for the optoelectronic response of the RGB sensor illuminated by the red, green or blue light; (b) The emission and relative response spectra of RGB LEDs at zero bias (red, green and blue from the top to bottom respectively).}\label{fig1}
\end{figure}

The OE response of the general LED array to the external optical field is very complicated, which may simultaneously contain the photoconductive and photovoltaic modes, and the process of the photocurrent generation caused by the dissociation of excitons. The LED receiver can be regarded as a current source when the impedance of the load is much smaller than its intrinsic impedance. Otherwise, different from the equivalent circuit of the LED receiver in Ref. \cite{20}, it should be slightly modified to a mixed model containing both the current source and the voltage source. In the case of load impedance $Z_0=1$ M$\Omega$ load, the DC response of the RGB IoT sensor can be approximately described by \beqa
I=\frac{V_R+V_G+V_B}{Z_t}=\frac{Z_RI_R+Z_GI_G+Z_BI_B}{Z_t},
\eeqa
where $I$ is the current over the load, which is simultaneously induced by the photovoltaic mode and the exciton dissociation in the LEDs.  $V_R$ ($I_R$), $V_G$ ($I_G$) and $V_B$ ($I_B$) represent the open-circuit voltages (short-circuit photocurrents) of the red, green, and blue LEDs respectively. $Z_R$, $Z_G$ and $Z_B$ are the effective reverse impedances of the red, green, and blue LEDs respectively, which depend on the power and spectrum of the injected light via the photoconductive mode. $Z_t=Z_R+Z_G+Z_B+Z_0$ is the total impedance in the sensor circuit. Usually, the reverse impedance of the LED receiver decreases with the photocurrent. In the linear response region, the short-circuit photocurrents and the open-circuit voltages of the individual R/G/B LEDs are given by
\beqa
I_{\mu}=\sum_{\nu=R,G,B}\eta_{\mu\nu}P_{\nu},~~~~
V_{\mu}=\sum_{\nu=R,G,B}Z_{\mu\nu}\eta_{\mu\nu}P_{\nu},
\eeqa
where $\eta_{\mu\nu}$ ($\mu,\nu\in\{R,G,B\}$) represents the responsivity of the $\mu$ color LED to the $\nu$ color light. $Z_{\mu\nu}$ represents the reverse impedance of the $\mu$ color LED receiver under $\nu$ color illumination. $P_R$, $P_G$ and $P_B$ are the optical powers of RGB light injected into the LED dice, respectively.

Parameters $Z_{\mu\nu}$ and $\eta_{\mu\nu}$ determine the characteristics of the LED sensor. Both the short-circuit current and the open-circuit voltage of the R/G/B LED receivers in the (R-R)/(B-G)/(B-B) links have been measured. It is found that $Z_{RR}$, $Z_{GB}$ and $Z_{BB}$ significantly decrease with the injected optical power, and $Z_{BB} \gg Z_{GB}$, $Z_{BB} \gg Z_{RR}$.
To experimentally investigate the characteristics of $\eta_{\mu\nu}$, we choose a single color LED as the receiver, and adjust the optical power of single color light in the RGB mixed illumination. The approximate linear responses exist over a wide input optical power range for red to red, red to green, green to blue, and blue to blue links. The responsivities are ordered as $\eta_{RR}\gg\eta_{GB}>\eta_{RG}>\eta_{BB}$, which is also consistent with the previous results \cite{12}.

\section{Results and analysis}

The DC optoelectronic responses of the RGB sensor to various monochrome lights or their combination are listed in Tab. \ref{tab1}, in which $P_r$ indicates the average optical power measured by the optical power meter placed at the same position and direction as the RGB sensor. The RGB light sources are powered alone or powered together, and the RGB components of the optical power are kept fixed. It is shown that the response voltage $V_r$ of the RGB sensor heavily depends on the spectral distribution of the light. Its photocurrent is not a trivial sum of the photocurrent of individual single color (red or green or blue) LED. This observation is different from the responsivity of the ordinary silicon-based photodiode, whose response follows the additivity of different spectral responses.

\begin{table}[h]
\centering
\caption{DC response of the RGB sensor with/without the occlusion}
\vspace{-0.3cm}
\begin{tabular}[t]{|c|c|c|c|c|c|c|c|c|}
\hline
{\scriptsize Transmitter} &{\scriptsize Dark} &{\scriptsize R} &{\scriptsize G} &{\scriptsize B} &{\scriptsize RG} &{\scriptsize RB} &{\scriptsize GB} &{\scriptsize RGB}\\
\hline
{\scriptsize $P_r$(mW)} &{\scriptsize 0.02} &{\scriptsize 0.53} &{\scriptsize 0.38} &{\scriptsize 0.64} &{\scriptsize 0.91} &{\scriptsize 1.16} &{\scriptsize 1.01} &{\scriptsize 1.54}\\
\hline
{\scriptsize $V_r$(mV)} &{\scriptsize 0.0} &{\scriptsize 0.0} &{\scriptsize 0.0} &{\scriptsize 25.1} &{\scriptsize 0.0} &{\scriptsize 89.1} &{\scriptsize 58.1} &{\scriptsize 91.0}\\
\hline
{\scriptsize $V_{r-R}$(mV)} &{\scriptsize 0.0} &{\scriptsize 0.0} &{\scriptsize 0.0} &{\scriptsize 24.6} &{\scriptsize 0.0} &{\scriptsize 24.5} &{\scriptsize 25.0} &{\scriptsize 25.0}\\
\hline
{\scriptsize $V_{r-G}$(mV)} &{\scriptsize 0.0} &{\scriptsize 0.0} &{\scriptsize 0.0} &{\scriptsize 0.0} &{\scriptsize 0.0} &{\scriptsize 0.0} &{\scriptsize 0.0} &{\scriptsize 0.0}\\
\hline
{\scriptsize $V_{r-B}$(mV)} &{\scriptsize 0.0} &{\scriptsize 0.0} &{\scriptsize 0.0} &{\scriptsize 23.6} &{\scriptsize 0.0} &{\scriptsize 88.0} &{\scriptsize 56.0} &{\scriptsize 89.0}\\
\hline
\end{tabular}\label{tab1}
\end{table}

From the table, we can also observe that the RGB sensor is not responsive to the DC red light or the DC green light or their mixture. The blue light is the key to excite the measurable DC photocurrent. Moreover, the DC red light and the DC green light play the incentive role in significantly enhancing the DC blue response. In this sense, it can be regarded as a light-programmable photodetector, i.e., utilizing the red or green light signal to program the responsivity of the blue light signal, or vice versa. This property may be intriguing in the simultaneous information transfer and wireless power transfer when phosphor-coated white LEDs or laser diodes (LD) are used as the transmitters.

The unique DC response behavior may also result from the dual mode interplay of the photovoltaic mode and photoconductive mode of the cascaded RGB LEDs \cite{24}. A partial occlusion experiment is conducted to investigate how the photovoltaic mode and the photoconductive mode operate. The results are also presented in Tab. \ref{tab1}, where $V_{r-\mu}$ means the response voltage when the $\mu$ color LED in the sensor is blocked (a minus sign). The results imply that the responses of the InGaN green LED to the blue light and the AlInGaP red LED to the red light are mainly due to the photovoltaic mode, while the responses of the AlInGaP red LED to the green or blue light tend to agree with the photoconductive mode. Blocking the green LED in the RGB sensor, any DC response will disappear no matter what the color of the injected light is, which means the response of the green LED to the blue light plays the most crucial role. When the green LED is in the darkness, it has a huge DC impedance. The blue LED only slightly contributes to the response, and the following analysis shows it has an anomalous impedance in the RGB sensor. The red LED has the highest optoelectronic responsivity to the red light. When the red LED is blocked, the total response approximately equals the response of the RGB sensor to the light without the red component.

The RGB sensor has no response to the DC optical signal with wavelength larger than 520 nm, like a photodetector with a short-pass filter. However, for response to AC optical signals, there does not exist such a forbidden band. Ordinarily, the photoconductive mode has the larger bandwidth than the photovoltaic mode. Due to the cascaded structure of the RGB sensor, there still exists the local reverse bias induced by neighbor LED's photocurrent. The dual modes of the photoconductive operation and photovoltaic operation may coexist in the RGB sensor. Its $3$-dB bandwidth of the frequency response to AC red light is at least $400$ kHz without the reverse bias and any equalization or impedance-match circuits. However its $3$-dB bandwidth to AC green light or blue light is much smaller than to the red light.

\vspace*{-0.2cm}
\begin{figure}[h]
\centering
\subfigure[]{\includegraphics[width=0.41\columnwidth,angle=-0]{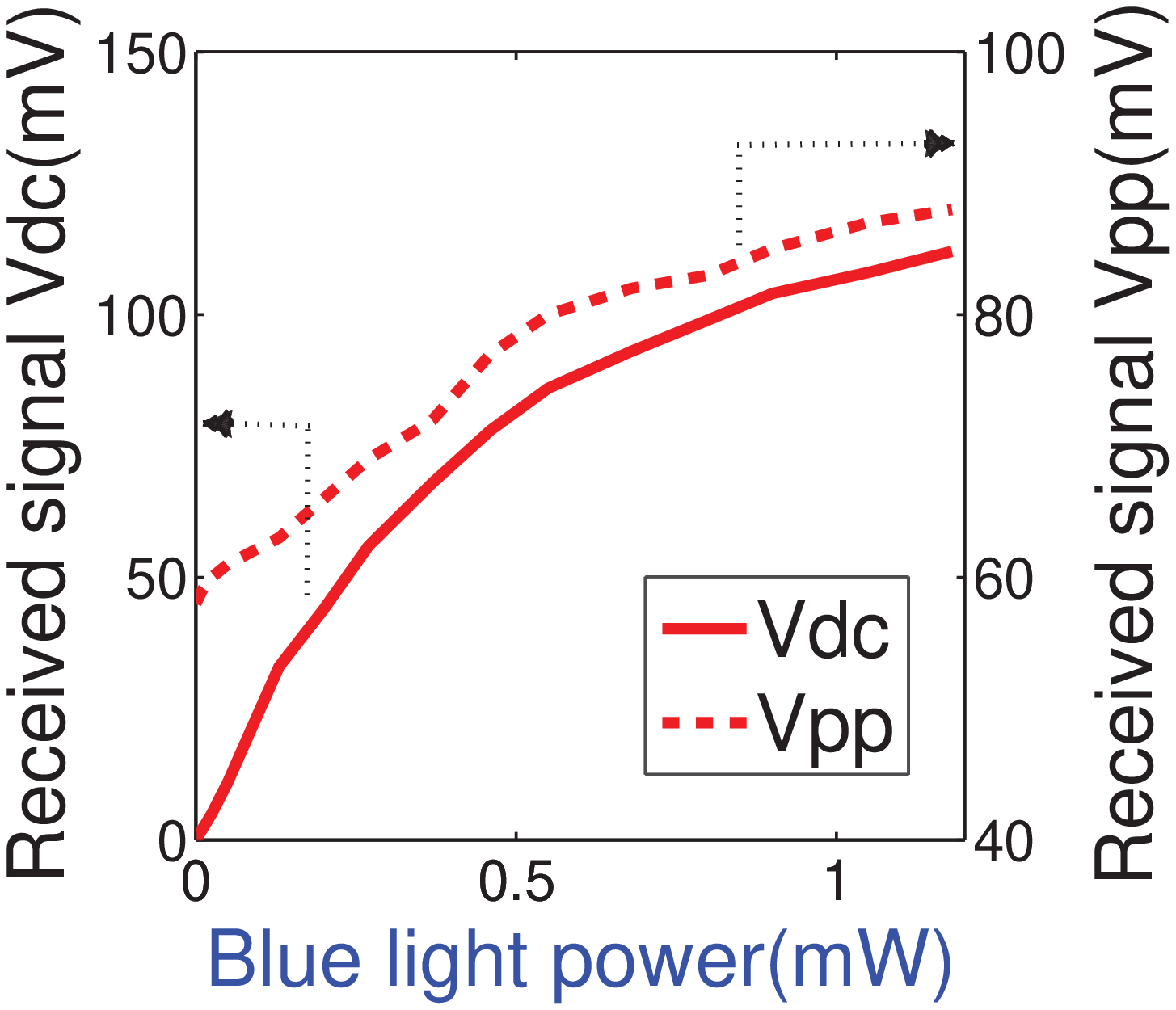}}
\subfigure[]{\includegraphics[width=0.41\columnwidth,angle=-0]{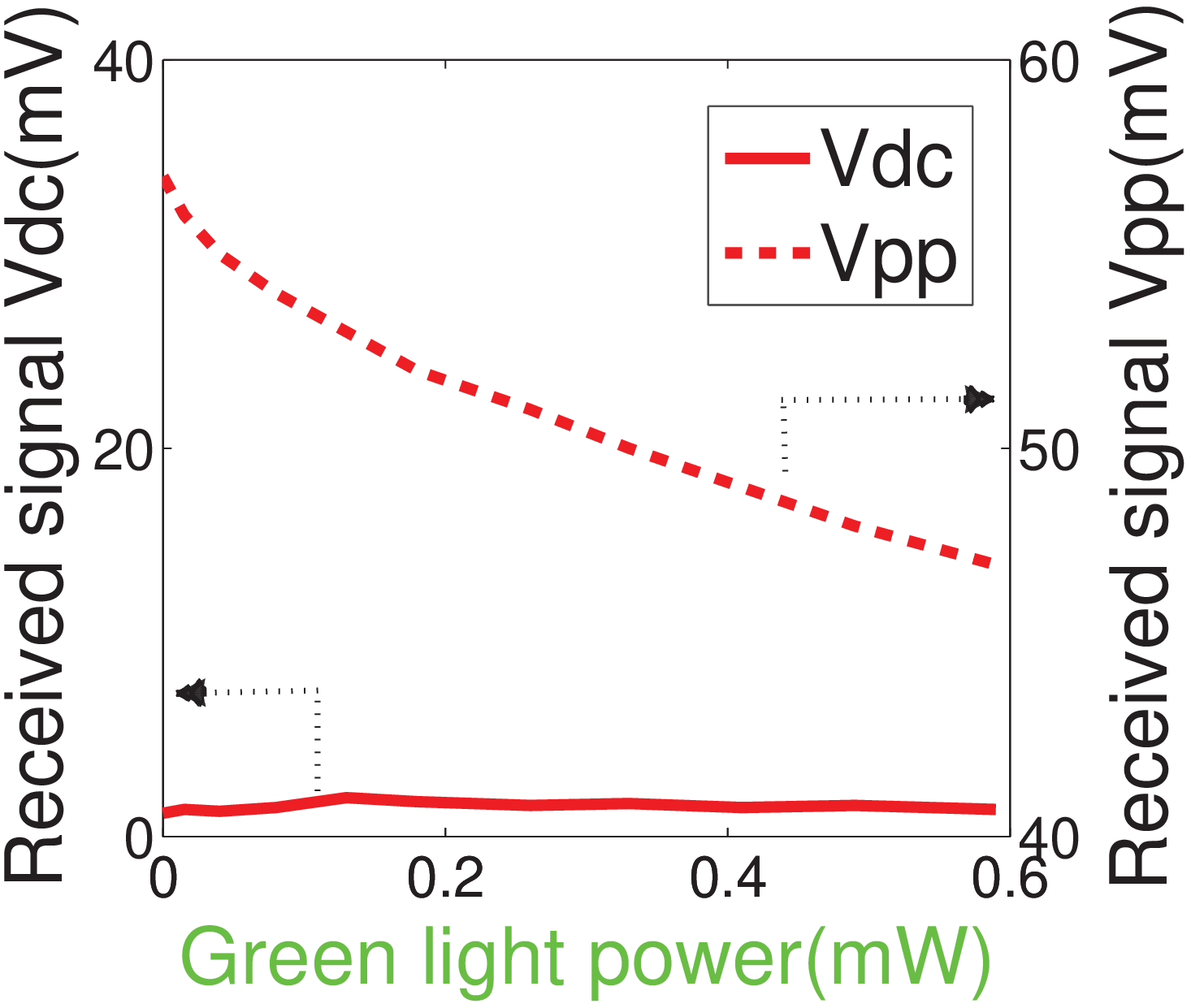}}
\subfigure[]{\includegraphics[width=0.41\columnwidth,angle=-0]{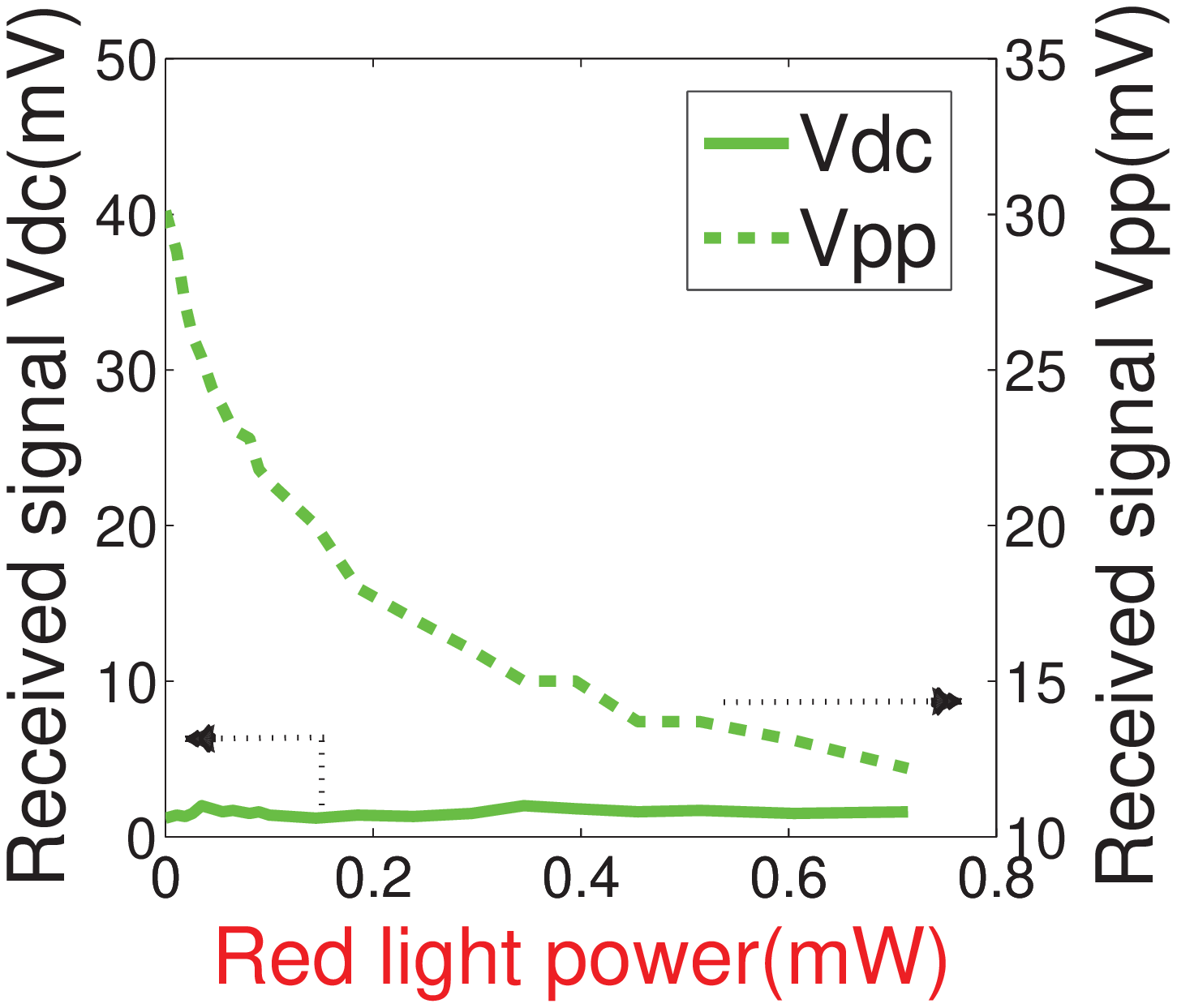}}
\subfigure[]{\includegraphics[width=0.41\columnwidth,angle=-0]{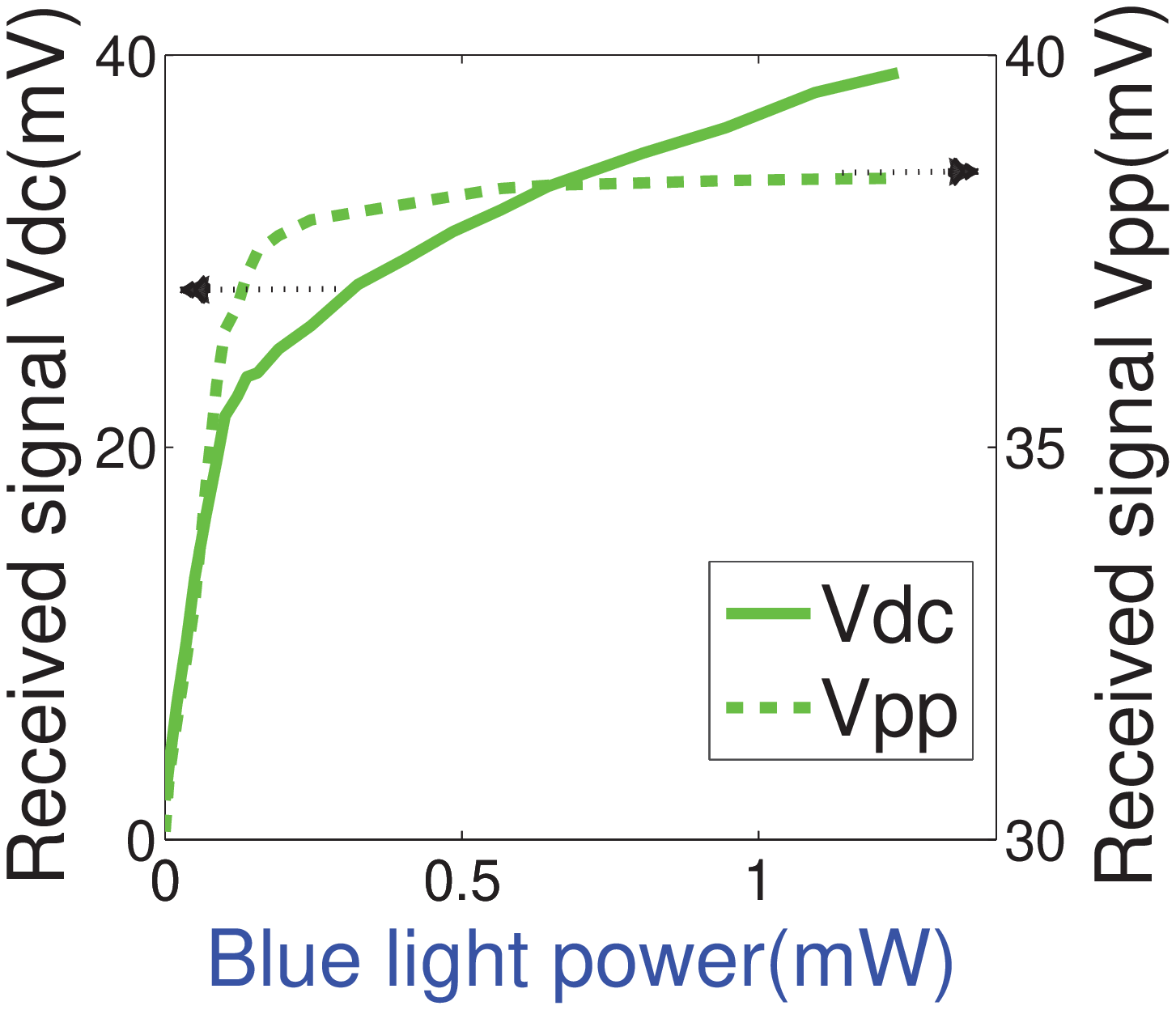}}
\subfigure[]{\includegraphics[width=0.41\columnwidth,angle=-0]{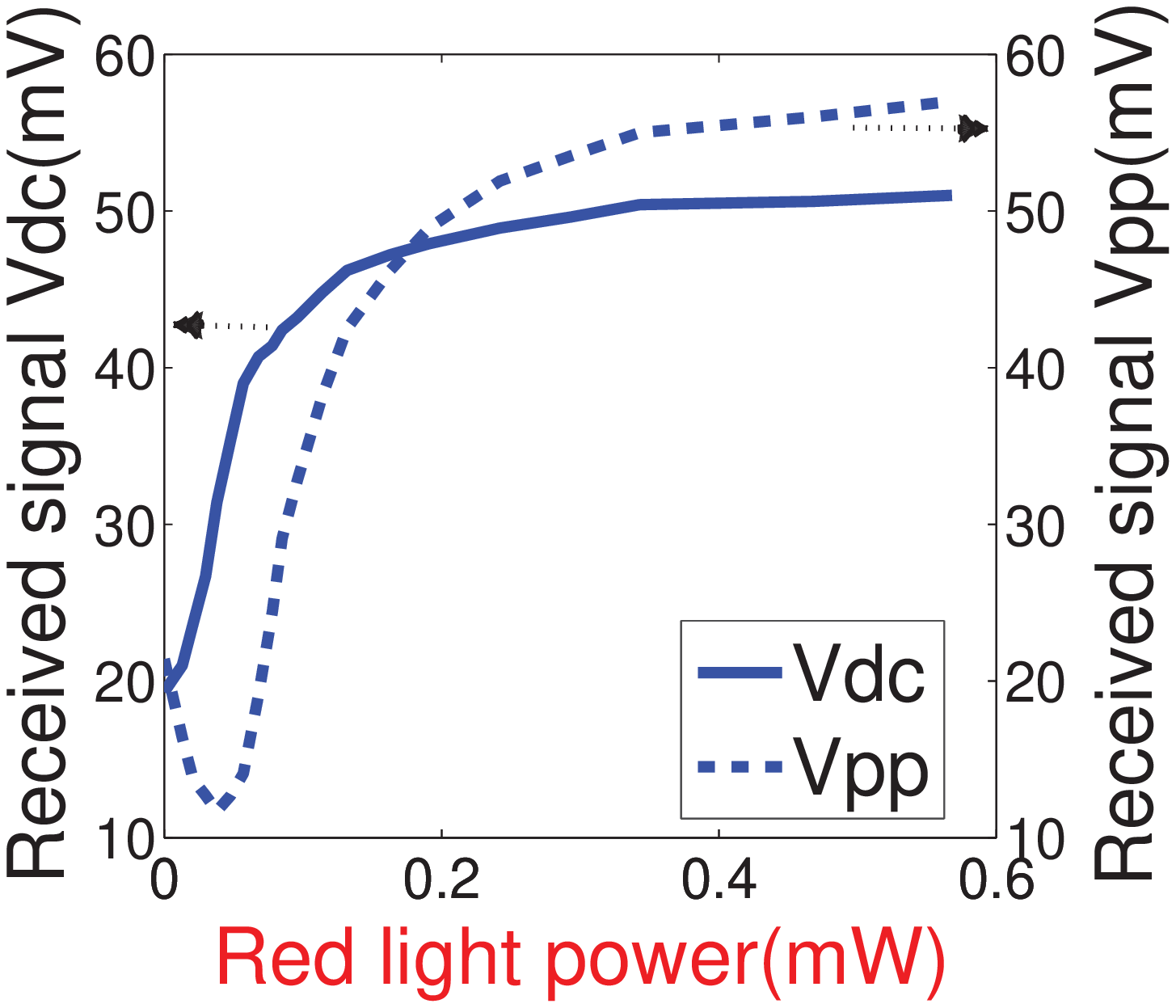}}
\subfigure[]{\includegraphics[width=0.41\columnwidth,angle=-0]{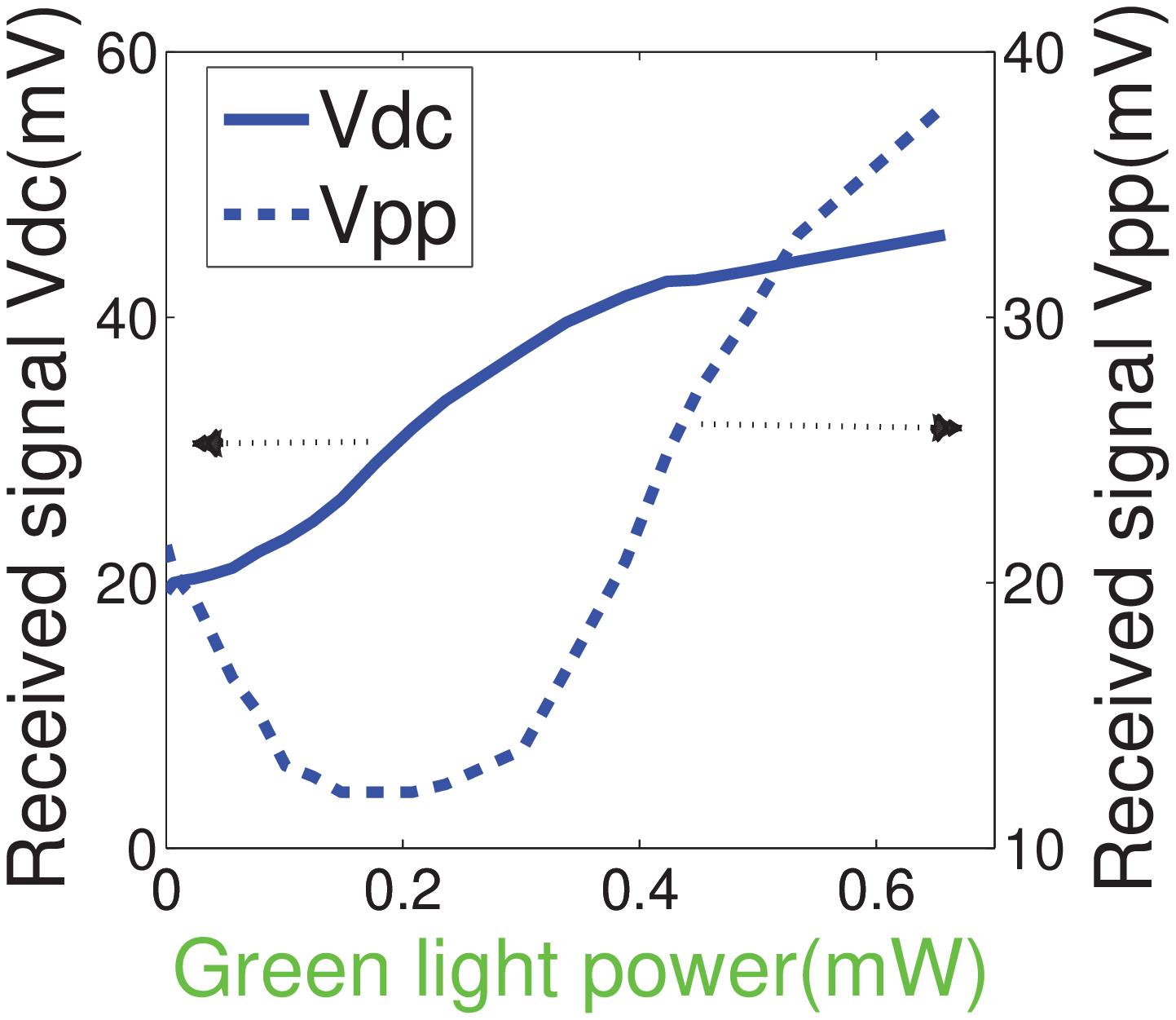}}
\caption{The light controllable DC/AC response of the RGB sensor to various combination of an input DC monochrome light and an input AC monochrome light: (a) DC blue light and AC red light at $f=1$ kHz, (b) DC green light and AC red light at $f=1$ kHz; (c) DC red light and AC green light at $f=500$ Hz, (d) DC blue light and AC green light at $f=500$ Hz; (e) DC red light and AC blue light at $f=100$ Hz, (f) DC green light and AC blue light at $f=100$ Hz.}
\label{fig2}
\end{figure}

The RGB sensor shows the light controllable response behaviors, depending on both DC and AC inputs and their colors. When a DC  monochrome color input and an AC monochrome color input at certain sinusoidal frequency $f$ are applied together, the sensor responses are plotted in Fig. \ref{fig2}. Every $1\times3$ series-connected LED in the transmitter has the same LED dice as the RGB sensor. The driving AC electrical signal of the LED transmitter is a positive-biased sinusoidal wave with DC voltage $V_{dc}=7.5$ V and peak-to-peak AC voltage $V_{pp}=3$ V. Therefore, the AC monochrome input is in fact a superposition of DC and AC signals of the same color. In each subplot of the figure, the curve color represents the input AC light color, the x-label color represents the input DC light color. The left and right y-axes represent the sensor responsive DC voltage and AC voltage respectively. Sinusoidal frequency for subplots (a)-(f) is accordingly set as {$1$ kHz, $1$ kHz, $500$ Hz, $500$ Hz, $100$ Hz, $100$ Hz.

As shown in Figs. \ref{fig2}(a) and (d), the DC blue light can significantly enhance the sensor response to red or green AC light input. The DC red or green light firstly suppresses and then enhances its response to blue AC light input as shown in Figs. \ref{fig2}(e) and (f). Interestingly, from Figs. \ref{fig2}(b) and (c), it is found that the DC green or red light suppresses its response to the red or green AC light input signal respectively. The mutual restraint of the green and red components of injected light can effectively decrease the interference in the VLC system using the phosphor-coated white LED transmitter (where communication signal is carried by the blue light) and the RGB sensor as the receiver.

\section{Conclusion}

The proposed RGB sensor consists of cascaded RGB LEDs. Based on experiments and analysis, the sensor shows similar properties as a multipole phototransistor and is easily controlled by variable number of color components of input light signals. The sensor can filter out the red and green DC optical signals, and naturally suppress the response to the red and green AC optical signals, but allows blue signals to pass through. Thus it is suitable for the VLC-based IoT applications and future programmable optical interconnections for the chips.

\newpage
\bibliographystyle{IEEEtran}

\end{document}